\documentclass[%
reprint,
amsmath,amssymb,
aps,
prb,
]{revtex4-1}

\usepackage{graphicx}
\usepackage{bm}
\usepackage{amsfonts}
\usepackage{amsthm}
\usepackage{amsbsy}
\usepackage{graphicx}
\usepackage{makeidx}
\usepackage{float}
\usepackage{xcolor}
\usepackage{epstopdf}
\newcommand*\diff{\mathop{}\!\mathrm{d}}

\DeclareMathOperator{\tr}{tr}

\newcommand{\stkout}[1]{\ifmmode\text{\sout{\ensuremath{#1}}}\else\sout{#1}\fi}

\begin{document}
\title{Thermal effects on a nonadiabatic spin-flip protocol of spin-orbit qubits}%
\author{Brecht Donvil}%
\email{brecht.donvil@helsinki.fi}
\affiliation{Department of Mathematics and Statistics, University of Helsinki, P.O. Box 68, 00014 Helsinki, Finland}
\author{Lara Ul\v{c}akar}%
\email{lara.ulcakar@ijs.si}
\affiliation{
Jozef Stefan Institute, Jamova 39, Ljubljana, Slovenia}
\affiliation{Faculty for Mathematics and Physics, University of Ljubljana, Jadranska 19, Ljubljana, Slovenia}
\author{Toma\v{z} Rejec}%
\email{tomaz.rejec@fmf.uni-lj.si}
\affiliation{
Jozef Stefan Institute, Jamova 39, Ljubljana, Slovenia}
\affiliation{Faculty for Mathematics and Physics, University of Ljubljana, Jadranska 19, Ljubljana, Slovenia}
\author{ Anton Ram\v{s}ak}%
\email{anton.ramsak@fmf.uni-lj.si}
\affiliation{
Jozef Stefan Institute, Jamova 39, Ljubljana, Slovenia}
\affiliation{Faculty for Mathematics and Physics, University of Ljubljana, Jadranska 19, Ljubljana, Slovenia}

\begin{abstract}
We study the influence of a thermal environment on a non-adiabatic spin-flip driving protocol of spin-orbit qubits. The driving protocol operates by moving the qubit, trapped in a harmonic potential, along a nanowire in the presence of a time-dependent spin-orbit interaction. We consider the harmonic degrees of freedom to be weakly coupled to a thermal bath. We find an analytical expression for the Floquet states and derive the Lindblad equation for a strongly non-adiabatically driven qubit. The Lindblad equation corrects the dynamics of an isolated qubit with Lamb shift terms and a dissipative behaviour. Using the Lindblad equation, the influence of a thermal environment on the spin-flip protocol is analysed.
\end{abstract}
\maketitle
\section{Introduction}

Electron-spin qubits are promising candidates as building blocks of quantum computers. They can be realized in gated semiconductor devices based on quantum dots and quantum wires \cite{wolf01,hanson07} and their state can be manipulated via magnetic fields  \cite{dresselhaus55,bychkov84} or the spin-orbit interaction, which is easily controlled with electrostatic gates \cite{stepanenko04,flindt06,coish06,sanjose08,golovach10,bednarek08,fan16,gomez12,pawlowski16,pawlowski16b,Pawlowski2017,Pawlowski2018}. Such systems were already experimentally realized in various semiconducting devices \cite{nadjperge12,nadjperge10,fasth05,fasth07,shin12}.

Recent studies proposed non-adiabatic protocols, where spin-qubit manipulation is achieved by translating a spin-qubit in one dimension \cite{cadez13,cadez14,Veszeli2018} in the presence of a time-dependent Rashba interaction \cite{nitta97,liang12,Yang2015}. Refs.~\cite{cadez13,cadez14} give exact analytical solutions for the time-dependent Schr\"{o}dinger equation of such a driven spin-qubit, confined in a harmonic potential, and express the spin rotation in terms of the non-adiabatic non-Abelian Anandan phase \cite{anandan88}. While qubit transformations in linear systems are limited to spin rotations around a fixed axis, this limitation can be eliminated on a ring structure \cite{kregar16,kregar16b}.

Manipulations of quantum systems are inevitably accompanied by external noise, coming from fluctuating electric fields {created} by the piezoelectric phonons \cite{sanjose08,sanjose06,huang13,echeveria13}, for example, or due to phonon-mediated instabilities in molecular systems with phonon-assisted potential barriers \cite{mravlje06,mravlje08}. Flying qubits could be carried by surface acoustic waves, where the noise can {arise} due to time dependence in the electron-electron interaction effects \cite{giavaras06,rejec00,jefferson06}. Recent related studies  \cite{Gefen05,lara17,lara18,Pyshkin2018, Li2018, Lu2018} considered the effects of additive noise present in the driving functions of the qubit. 

In order to study effects of a thermal environment on the qubit manipulation, we aim to derive an effective dynamics for the spin-qubit from a full microscopic model of the qubit weakly interacting with a thermal bath.
In the case of weak interactions, there exists a well-known approximation scheme to integrate out the dynamics of the bath and obtain the effective dynamics for the system, which is given by the Lindblad equation \cite{BreuerBook}. For adiabatic and weakly non-adiabatic driving of the system, the aforementioned weak-coupling scheme still holds and leads to a slightly modified Lindblad equation. However, for strongly non-adiabatic driving, which we consider in the present work, the necessary assumptions for this approximation scheme break down.

In a recent work\cite{Dann2018}, the authors showed how a modified weak-coupling scheme can be performed in order to derive the Lindblad equation for an arbitrarily driven weakly-coupled system. In the case of periodic driving, this result reduces to the earlier derived Floquet-Lindblad form \cite{BlBu1991,BPfloquet}. Although these results show how in principle a Lindblad equation can be derived, employing these methods essentially requires one to solve the evolution of the driven system in absence of the bath. In general, obtaining an analytic expression for the evolution of a driven system is non trivial and one has to resort to numerical approximations. In the present work, however, we consider periodic driving for which the analytic solution is known \cite{cadez13}, giving us access to the exact Floquet-Lindblad equation.

The paper is structured as follows: in Section~\ref{sec:model} we introduce the model and the coupling to the thermal bath. In Section~\ref{sec:Floquet} the Floquet formalism and the corresponding exact solutions are given, which, in Section~\ref{sec:Lind}, serves as a basis for an exact derivation of the Lindblad operators and the Floquet-Lindblad equation. 
The formalism is then applied to a simple example of a non-adiabatic driving and spin rotation in Section~\ref{sec:example}. In Section~\ref{sec:conclusion} we give conclusions and in Appendices~A and B we present derivations of individual terms of the Floquet-Lindblad equation.

\section{Model}\label{sec:model}
Our system of interest is a spin-qubit represented as an electron  confined in a quantum wire with a harmonic potential.\cite{cadez13,cadez14} The centre of the trap,  $\xi(t)$, can be arbitrarily translated along the wire by means of time-dependent external electric fields. 

The spin-orbit Rashba interaction couples the electron's spin with its orbital motion, resulting in the system Hamiltonian 
\begin{equation}\label{H}
H(t)=\frac{p^{2}}{2m^{*}}+\frac{m^{*}\omega_0^{2}}{2}\big(x-\xi(t)\big)^{2}+\alpha(t)\,p\, \sigma_y,
\end{equation}
where $m^{*}$ is the effective electron mass, $\omega_0$ is the frequency
of the harmonic trap, and $p$ and $\sigma_y$ are the momentum and spin operators, respectively. The strength of the spin-orbit interaction $\alpha(t)$ is time 
dependent due to time dependent external electric fields and the spin rotation axis is fixed along the $y$-direction \cite{nadjperge12}.  Throughout the paper we set $\hbar=1$ and initial time $t=0$. The exact time dependent  solution of the Schr{\"o}dinger equation corresponding to the Hamiltonian Eq.~\eqref{H} is given by\cite{cadez14} 
\begin{equation}\label{eq:psi}
|\Psi_{}(t)\rangle=U(t,0)|\Psi_{}(0)\rangle, 
\end{equation}
where, in the time evolution operator 
\begin{equation}\label{eq:ev}
U(t,0)=\mathcal{U}^\dagger(t)e^{-i H_{0} t}\mathcal{U}(0),
\end{equation}
$H_0$ represents the time independent harmonic oscillator, i.e.,  Eq.~(\ref{H}) with $\xi(t)=\alpha(t)=0$, and 
\begin{subequations}
\begin{eqnarray}\label{psi}\label{eq:transf}
\mathcal{U}^\dagger(t)&=&e^{-{i }(\varphi_0(t)+\varphi(t) \sigma_y)}\mathcal{A}_{\alpha}(t)\mathcal{X}_{\xi}(t),\\
\mathcal{A}_{\alpha}(t)&=&e^{-i\dot{a}_{c}(t)p\sigma_y/\omega_0^2}
e^{-im^{*}a_{c}(t){x}\sigma_y},\\
\mathcal{X}_{\xi}(t)&=&e^{im^{*}\big(x-x_{c}(t)\big)\dot{x}_{c}(t)}e^{-ix_{c}(t)p}.
\end{eqnarray}
\end{subequations}
The unitary transformations $\mathcal{X}_{\xi}(t)$ and $\mathcal{A}_{\alpha}(t)$ are completely determined by the classical responses $x_c(t)$ and $a_c(t)$ to the driving. The responses solve the differential equations
\begin{subequations}\label{xcac}
\begin{eqnarray}
\ddot{x}_c(t)+\omega_0^{2}x_{c}(t)&=&\omega_0^{2}\xi(t),\\
\ddot{a}_c(t)+\omega_0^{2}a_{c}(t)&=&\omega_0^{2}\alpha(t).
\end{eqnarray}
\end{subequations}
$\varphi(t)=-m^{*}\int_{0}^{t}\dot{a}_{c}(\tau)\xi(\tau)\mathrm{d}\tau$ while the time dependent phase $\varphi_0(t)$, given in Ref.~\onlinecite{cadez14}, is irrelevant for the time evolution operator Eq.~(\ref{eq:ev}) and will be omitted. Also note that the time evolution operator is invariant with respect to the gauge transformation 
\begin{equation}\label{gauge}
\mathcal{U}^\dagger(t) \to e^{i(\delta_1+\delta_2 \sigma_y)} \mathcal{U}^\dagger(t),
\end{equation} 
where $\delta_1$ and $\delta_2$ are real constants.

The system described by the time dependent $H(t)$ is coupled to a bosonic thermal bath. The total Hamiltonian of the spin-qubit interacting with the bath is
\begin{equation}\label{eq:Hamiltonian}
H_{\rm tot}(t)=H(t)+H_B+H_I(t),
\end{equation}
where the bath Hamiltonian $H_B$ represents a set of oscillators
\begin{equation}
H_B=\sum_{\mathbf{k}} \omega_{\mathbf{k}}b^\dagger_{\mathbf{k}} b_{\mathbf{k}},
\end{equation}
where $b^\dagger_{\mathbf{k}}$ ($ b_{\mathbf{k}}$) are creation (annihilation) operators and the oscillators have a linear dispersion relation $\omega_{\mathbf{k}}=c|\mathbf{k}|$, $\mathbf{k}\in \mathbb{R}^3$. We consider only states with energies below a cut-off energy $\omega_c$.

The spin-qubit is coupled to the bath through the interaction Hamiltonian $H_I(t)\propto\left(x-\xi(t)\right)\sum_{\mathbf{k}} x_{\mathbf{k}}$ which couples the respective position operators, 
\begin{equation}\label{eq:intHam}
H_I(t)=g\sqrt{\frac{\pi c^3}{\omega_0 v}}\left(a^\dagger+a-\sqrt{2m^*\omega_0}\xi(t)\right)\sum_{\mathbf{k}}(b_{\mathbf{k}}+b^\dagger_{\mathbf{k}}),
\end{equation}
where $g$ is a dimensionless coupling strength, $a^\dagger$ ($ a$) is a bosonic creation (annihilation) operator of the harmonic trap and $v$ is the volume of the bath.

\section{Floquet Theory for quantum systems}\label{sec:Floquet}
Before deriving the Lindblad equation for the driven spin-qubit, it is instructive to briefly discuss the Floquet theory for periodically driven quantum systems. For systems with a Hamiltonian periodic in time, $H(t+T_d)=H(t)$, it is possible to describe the time-evolution in terms of periodic eigenvectors of the Schr\"{o}dinger equation called the Floquet states \cite{Shirley,Zeldovich}. The Floquet states $|\phi_q(t)\rangle$ form a complete basis and are defined as solutions of the eigenvalue problem
\begin{subequations}\label{eq:Flset}
\begin{align}
&(H(t)-i\partial_t)|\phi_q(t)\rangle=\epsilon_q|\phi_q(t)\rangle,\\
&|\phi_q(t+T_d)\rangle=|\phi_q(t)\rangle,
\end{align}
\end{subequations}
where $\epsilon_q$ is called a quasi-energy. The evolution of an arbitrary initial state $|\psi_0\rangle$ from an initial time $0$ to time $t$, expressed in terms of the Floquet states, is
\begin{equation}
|\psi(t)\rangle=\sum_q e^{-i\epsilon_q t}|\phi_q(t)\rangle \langle\phi(0)|\psi_0\rangle.
\end{equation}

In practice, solving Eq.~\eqref{eq:Flset} proves to be non-trivial. However, with a gauge transformation $V(t)$ such that
\begin{equation}\label{eq:flfl}
G=V(t)H(t)V(t)^\dagger+i(\partial_t V(t))V(t)^\dagger
\end{equation} 
is time independent, the Floquet states and quasi-energies can be found in terms of eigenvectors and eigenvalues of $G$. Let $|q\rangle$ be an eigenvector of $G$ with the eigenvalue $\epsilon_q$. One can check that the state 
\begin{equation}\label{eq:fl}
|\phi_q(t)\rangle= V(t)^\dagger |q\rangle
\end{equation}
is a Floquet state with the quasi-energy $\epsilon_q$. Note that $\mathcal{U}(t)$, as defined in Eq.~\eqref{eq:transf}, has exactly the property Eq.~\eqref{eq:flfl} with $G=H_0$. Therefore, the Floquet states of the driven qubit are $|\phi_q(t)\rangle= \mathcal{U}^\dagger(t) |\psi_q\rangle$, where $|\psi_q\rangle$ is an eigenstate of $H_0$ with the energy $\epsilon_q=\omega_0(q+\frac{1}{2})$.

\section{Derivation of the Lindblad equation}\label{sec:Lind}
The reduced density matrix of the spin-qubit at time $t$ in the Schr\"{o}dinger picture is 
\begin{equation}
\bar{\rho}(t)=\tr_B\left( U_{\textrm{tot}}(t,0)\rho_{\textrm{tot}}(0)U_{\textrm{tot}}^\dagger(t,0)\right)
\end{equation}
where $\mathrm{tr}_B$ denotes the trace over bath degrees of freedom, $U_\textrm{tot}(t,0)$ the time evolution operator of the whole system, and
\begin{equation}
\rho_\textrm{tot}(0)=\rho(0)\otimes \frac{e^{-\beta H_B}}{\tr_B e^{-\beta H_B}},
\label{rhotot}
\end{equation}
is an initially separable density matrix consisting of the qubit in the state $\rho(0)$ and the bath at the inverse temperature $\beta=\frac{1}{k_BT}$. In the interaction picture,
\begin{equation}
\rho(t)=U^\dagger(t,0)\bar{\rho}(t)U(t,0)
\end{equation} 
where $U(t,0)$ is the time evolution operator of the qubit, the Floquet-Lindblad equation for the qubit interacting with the bath via Eq.~\eqref{eq:intHam} is of the form\cite{BreuerBook}
\begin{align}\label{eq:lind}
\frac{\diff}{\diff t}\rho(t)=-{i}\big[H_{LS},\rho(t)\big]+\mathcal{D}\big(\rho(t)\big).
\end{align}

In what follows we assume that the driving frequency $\omega_d=\frac{2\pi}{T_d}$ is $\omega_d=\frac{\omega_0}{n_d}$ with $n_d\in\mathbb{N}$, as appropriate for the spin-flip protocol studied in this paper.
The first term on the right hand side of Eq.~(\ref{eq:lind}) contains the Lamb shift Hamiltonian
\begin{equation}\label{Lambshift}
H_{LS}=\sum_{n\in \mathbb{Z}} S(n\omega_d) A_n^\dagger A_n,
\end{equation}
where $A_n$ are the Lindblad operators (to be defined below) and 
\begin{equation}
S(\omega)=\frac{g^2}{2\pi\omega_0}{\cal P}\!\!\int_0^{\omega_c} \diff \omega' \omega'^2\left(\frac{1+N(\omega')}{\omega-\omega'}+\frac{N(\omega')}{\omega+\omega'}\right)
\end{equation}
with $\cal P $ denoting the principal value and
\begin{equation}
N(\omega)=\frac{1}{\exp (\beta\omega)-1}
\end{equation} 
being the Bose occupation numbers of the bath degrees of freedom. Throughout the paper we express the temperature in terms of the average occupation of the system oscillator $n_T=N(\omega_0)$. In Appendix~\ref{secapp:ladder1} we show the Lamb shift Hamiltonian in an explicit form.

The second term on the right hand side of Eq.~\eqref{eq:lind} is the dissipator term
\begin{equation}\label{eq:diss}
\mathcal{D}\big(\rho(t)\big)=\sum_{n\in \mathbb{Z}}  \gamma(n\omega_d)\left(A_n\rho(t)A_n^\dagger -\tfrac{1}{2}\{A_n^\dagger A_n,\rho(t)\}\right),
\end{equation}
with rates
\begin{equation}
\gamma(\omega)=\begin{cases}g^2\frac{\omega^2}{\omega_0}(1+N(\omega)),  &\omega\ge0,\\g^2\frac{\omega^2}{\omega_0}N(|\omega|),&\omega<0.\end{cases}
\end{equation}

The Lindblad operators are obtained by finding $A_n$ such that \cite{BPfloquet,BreuerBook}
\begin{eqnarray}\label{eq:ladders}
U^\dagger(t,0)\left(a^\dagger+a-\sqrt{2m^*\omega_0}\xi(t)\right)U(t,0)\nonumber\\
=\sum_{n\in\mathbb{Z}} A_n e^{-i n\omega_d t}.
\end{eqnarray}
We present the actual calculation of the Lindblad operators in Appendix~\ref{secapp:ladder}. The result for $n\ge0$ is
\begin{eqnarray}\label{eq:ladder1}
A_n&=&\delta_{n,n_d}\,a\nonumber
+\sqrt{{ 2m^* \omega_0}}\bigg(\hat{x}_{c,n}-\hat{\xi}_n+\frac{\hat{\dot{a}}_{c,n}}{\omega^2_0}\sigma_y+
 \nonumber\\
& &+\delta_{n,n_d}\frac{1}{2}\left(-\frac{\dot{a}_c(0)}{\omega_0^2}+i\frac{a_c(0)}{\omega_0}\right)\sigma_y+
\nonumber\\
& &+\delta_{n,n_d}\frac{1}{2}\left(-x_c(0)-i\frac{\dot{x}_c(0)}{\omega_0}\right) \bigg),
\end{eqnarray}
and $A_{-n}=A_n^\dagger$. Here by $\hat{f}_n$ we denote a Fourier component of the function $f$. The Lindblad operators are completely determined by the solutions of the response Eqs.~\eqref{xcac}. 

We can greatly simplify the form of the dissipator Eq.~\eqref{eq:diss} by absorbing the constant terms in the Lindblad operators into the Lamb shift. This procedure is outlined in Appendix~\ref{secapp:ladder1}. Let us define a rate
\begin{equation}
\bar{\gamma}=\frac{2 m^* }{\omega_0^3}\sum_{\substack{n\in \mathbb{Z}\\ n\ne\pm n_d}} \gamma(n\omega_d)|\hat{\dot{a}}_{c,n}|^2,
\end{equation}
and a new Lindblad operator
\begin{equation}
\bar{A}=a+\sqrt{\frac{m^* }{2\omega_0}}\bigg(2\frac{\hat{\dot{a}}_{c,n_d}}{\omega_0}-\frac{\dot{a}_c(0)}{\omega_0}+ia_c(0)\bigg)\sigma_y.
\end{equation}
With these definitions, Eq.~\eqref{eq:lind} reduces to a much simpler form
\begin{align}\label{eq:Lindf}
\frac{\diff}{\diff t}\rho(t)=&-{i}\left[\bar{H}_{LS},\rho(t)\right]+\bar{\gamma} \left(\sigma_y\rho(t)\sigma_y-\rho(t)\right)+
\nonumber\\&+ \gamma(\omega_0)\left(\bar{A}\rho(t)\bar{A}^\dagger-\tfrac{1}{2}\left\{\bar{A}^\dagger\bar{A},\rho(t)\right\}\right)+\nonumber\\&+ \gamma(-\omega_0)\left(\bar{A}^\dagger\rho(t)\bar{A}-\tfrac{1}{2}\left\{\bar{A}\bar{A}^\dagger,\rho(t)\right\}\right).
\end{align}
The redefined Lamb shift Hamiltonian $\bar{H}_{LS}$ is shown explicitly in Appendix~\ref{secapp:ladder1}.
The dissipator in the above equation consists of two types of terms: the term proportional to $\bar{\gamma}$ is a dephasing term and causes a decay of spin size. The other terms lead to thermal activation in the oscillator component.

\section{Example}\label{sec:example}

At the initial time $t=0$, let the electron be in the ground state manifold of $H(0)$, spanned by a Kramers doublet. In particular, we choose the qubit to be in the spin-up state, i.e., with $\langle\Psi(0)| \sigma_x |\Psi(0)\rangle=\langle\Psi(0)| \sigma_y |\Psi(0)\rangle=0$. Using Eq.~(\ref{gauge}), such a state can be constructed as
\begin{equation}
|\Psi(0)\rangle=e^{i m^*\big({a}_{c}(0){x}_{c}(0)+\frac{\dot{a}_{c}(0)\dot{x}_{c}(0)}{\omega_0^2}\big)\sigma_y}
\mathcal{U}^\dagger(0)|\psi_{0}\rangle|\chi_{\uparrow}\rangle.
\end{equation}
Here $|\psi_{0}\rangle$ is the ground state of the harmonic oscillator Hamiltonian $H_0$ and $|\chi_{\uparrow}\rangle$ is the up-state spinor in the eigenbasis of $\sigma_z$. The initial density matrix of the qubit Eq.~(\ref{rhotot}) is $\rho(0)=|\Psi(0)\rangle\langle\Psi(0)|$.

As a simple example of the theory outlined in the previous section, we consider periodic spin transformations following an elliptic path in the parametric space $[\xi(t),\alpha(t)]$ with 
\begin{subequations}\label{driving}
	\begin{align}
		&\xi(t) = \xi_0 \cos\frac{\omega_0 t}{2},\\
		&\alpha(t)=\alpha_0-\alpha_0 \sin\frac{\omega_0 t}{2},
	\end{align}
\end{subequations}
i.e, $n_d=2$. This choice of driving, together with initial conditions $x_c(0)=\xi(0)$, $a_c(0)=\alpha(0)$ and  $\dot{x}_c(0)=0$, $\dot{a}_c(0)=0$ for differential equations \eqref{xcac}, leads to classical responses 
\begin{subequations}\label{eq:exresp}
	\begin{eqnarray}
	x_c(t)&=&\frac{\xi_0}{3} \left(4   \cos\frac{ \omega_0 t}{2}-\cos\omega_0 t \right),\\
	a_c(t)&=&\alpha_0-\frac{\alpha_0}{3}  \left(4 \sin\frac{\omega_0 t}{2} -2\sin \omega_0 t\right).
	\label{response}
	\end{eqnarray}
\end{subequations}
The driving guarantees that after a completed cycle the state $|\Psi(T_d)\rangle$
returns to the ground state -- with spin rotated around the $y$-axis by the Anandan quantum phase $\phi_A=2\varphi(T_d)$ determined solely by the contour $\mathcal{C}_{\xi}$ in the parametric space $[\xi(t),a_c(t)]$ or, equivalently,  $\mathcal{C}_{\alpha}$ in the space $[x_c(t),\alpha(t)]$,
\begin{equation}\label{anandan}
\phi_A=2m^{*}\oint_{\mathcal{C}_{\xi}} a_{c}[\xi]{\rm d}\xi=2m^{*}\oint_{\mathcal{C}_{\alpha}}x_{c}[\alpha]{\rm d}\alpha.
\end{equation}
The corresponding contour ${\cal{C}}_{\alpha}$ is shown in Fig.~\ref{fig:Bloch}(a) with a full black line and the dashed line indicates the driving protocol $[\xi(t),\alpha(t)]$.  Note that the area of the shaded region equals $\phi_A/(2m^*)$.

During the motion of the system, the electron's spatial wave function is a superposition containing also the oscillator's excited states and the
spin of the electron is rotated around the $y$-axis by the angle $\phi(t)$. After a completed driving cycle the electron's spin has, according to Eq.~(\ref{anandan}), rotated by the angle $\phi(T_d)=\phi_A =\frac{8}{3} \pi m^*\alpha_0\xi_0$.  

In the upcoming numerical studies we take $\gamma/\omega_0\lesssim \frac{1}{10}$. This is to ensure that the effect of the bath remains a perturbation to the free dynamics of the spin-qubit, which is the underlying assumption for the Lindblad equation Eq.~\eqref{eq:lind}. We also fix 
the Rashba coupling $\alpha_0=\frac{3}{16} (\omega_0/m^*)^{1/2}$ and $\xi_0=(m^*\omega_0)^{-1/2}$ such that $\phi_A={\pi \over 2}$, i.e., two driving cycles are needed for a spin-flip. The cut-off frequency is set to $\omega_c=2 \omega_0$ and all numerical calculations were carried out using the QuTiP framework \cite{qutip12,qutip13}.

\subsection{Spatial position and spin properties}
\begin{figure}[h]
	\centerline{\includegraphics[width=5 cm]{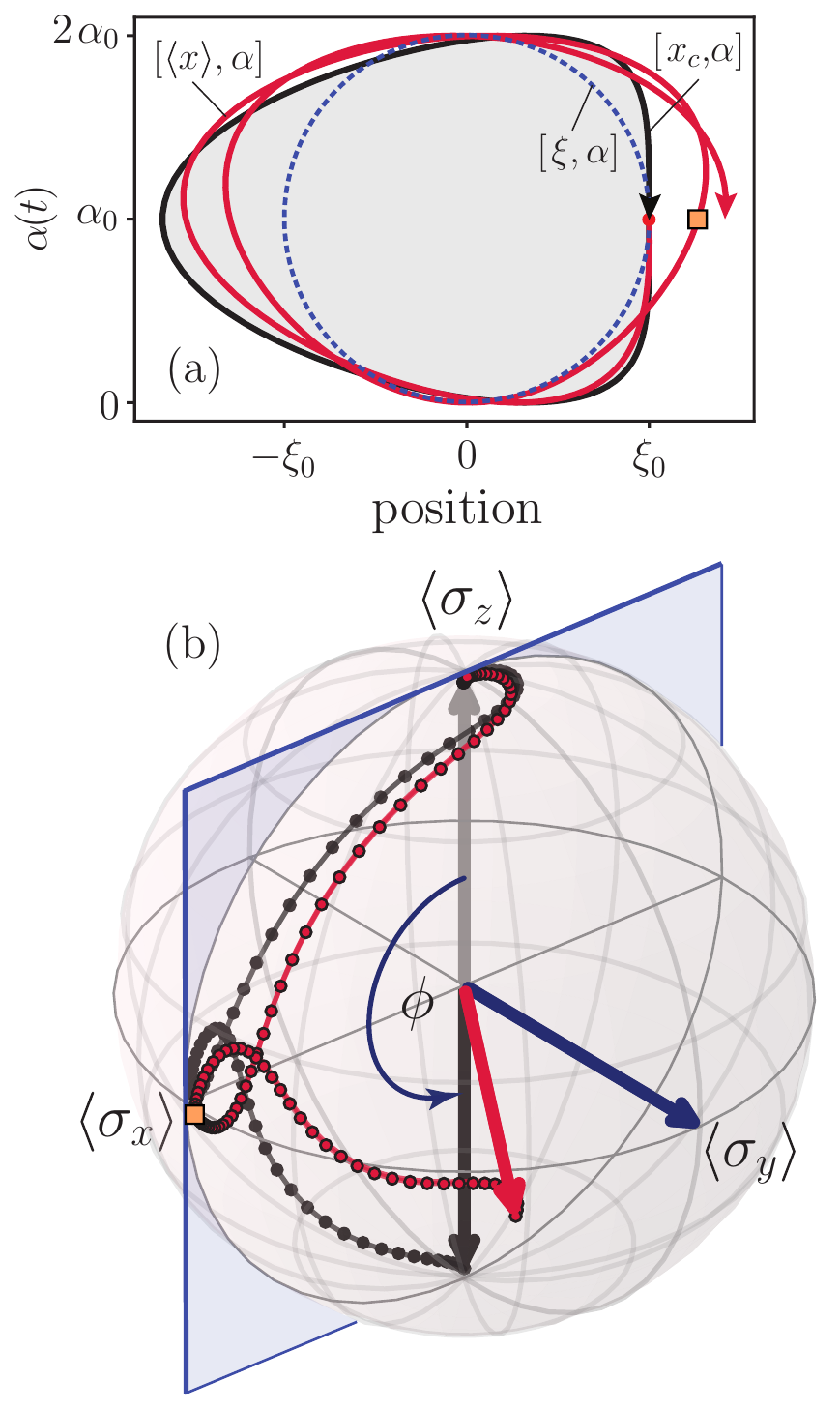}}
	\caption{(a) Contour ${\cal{C}}_{\alpha}$  of the driving and the response in the parametric space spanned by the spatial position and the spin-orbit driving. The red bullet represents the starting point and the orange square the response $[\langle x\rangle,\alpha]$ after one cycle for $g=0.3$, $n_T=1$ and $\omega_c=2\omega_0$. (b) Two cycles of the qubit rotation within the Bloch sphere: the black bullets and the black arrow represent the noninteracting result, $g=0$, with spin rotation around the $y$-axis for an angle $\phi=\pi$. The red bullets and the red arrow show the rotation for $g=0.2$, $n_T=1$ and $\omega_c=2\omega_0$. Bullets show positions at equal time intervals.}
	\label{fig:Bloch}
\end{figure}

In Fig.~\ref{fig:Bloch}(b) we show the time dependence of the expectation value of the spin. Since we are considering rotations around the $y$-axis, $\langle \sigma_y \rangle=0$, i.e., the expectation value is confined to the $x$-$z$ plane and {\it within} the Bloch sphere.  Bullets represent the values at equal time steps for the total time duration of two cycles. Note that in the absence of the interaction, $g=0$, the spin transformation is exactly one spin-flip (black). The red bullets represent the corresponding result for $g=0.2$, $n_T=1$ and $\omega_c=2\omega_0$. The orange square indicates the spin at $t=T_d$.

The interaction with the bath influences the spin-flip protocol, namely: the angle of the spin rotation, the size of the spin, the expectation value of the position  $\langle x \rangle$ and the oscillator part of the wave function, which does not return to the ground state at the end of the transformation. Below we discuss these effects.

\begin{figure}[h]
	\centerline{\includegraphics[width=6.5 cm]{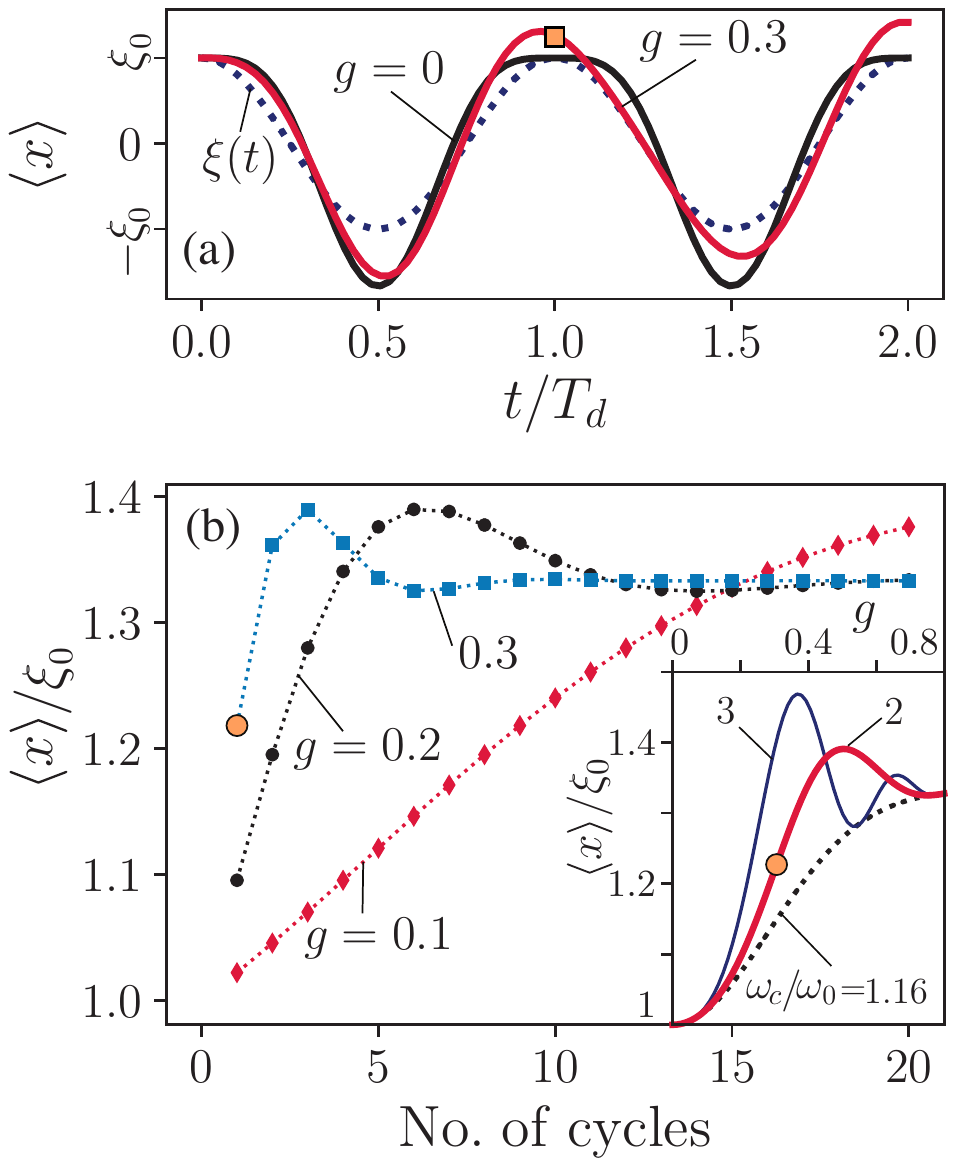}}
	\caption{(a) Driving $\xi(t)$ as a function of time for two cycles (dotted line) and the expected position of the electron $\langle x \rangle$ in the case of no interaction (black) and for $g=0.3$, $n_T=1$ (red). The orange square corresponds to the square in Fig.~\ref{fig:Bloch}. (b) $\langle x \rangle$ as a function of the number of qubit transformation cycles, for different values of the coupling $g$ at $T=0$. Inset:  $\langle x \rangle$ after one cycle as a function of $g$ for $\omega_c/\omega_0=1.16$, 2, and 3. The orange circle corresponds to $g=0.3$.} 	\label{fig:x}
\end{figure}

The influence of the interaction with the bath on the expected position of the oscillator $\langle x\rangle$ can be observed from Figs.~\ref{fig:Bloch}(a) and \ref{fig:x}(a) by comparing the result for the non-interacting pure dynamics (black line) to the interacting result (red line). In the non-interacting case, the expected position is equal to the classical response, $\langle x\rangle=x_c(t)$, as can easily  be shown from Eq.~\eqref{eq:transf}. In the presence of the bath $\langle x\rangle$ deviates from $x_c(t)$ and is shifted from the starting point (red bullet) after one cycle (orange square). Fig.~\ref{fig:x}(a) shows that the shift of $\langle x\rangle$ to larger values is increased after the second cycle, $t=2T_d$, and is further increased until the driven system reaches a steady state. This is in agreement with the classical result for a driven weakly damped harmonic oscillator, $x_c \to {4 \over 3}\xi_0$, after a large number of cycles [the first term in Eq.~(\ref{eq:exresp}a)]. 
The approach to the steady state can be observed in Fig.~\ref{fig:x}(b) which shows $\langle x\rangle$  in terms of the number of driving cycles (needed to half-flip the spin), for different values of the coupling strength $g$ at $T=0$. Due to the low frequencies introduced by the Lamb shift in Eq.~\eqref{eq:Lindf}, $\langle x\rangle$  exhibits slow damped oscillations. 
The inset of Fig.~\ref{fig:x}(b) shows $\langle x\rangle$ after one cycle as a function of $g$ for various values of the cut-off energy $\omega_c$. Note the absence of oscillations for  $\omega_c=1.16\, \omega_0$, dotted line. At this cut-off energy $S(\omega_0)+S(-\omega_0)\to 0$ and consequently spatial parts of $H_{LS}\to 0$, see Eqs.~(\ref{HLS}) and (\ref{Spm}), which leads to results qualitatively close to the classical result for a damped oscillator. At higher cut-off energies quantum terms in $H_{LS}$ leads to a much richer dynamics. 
The orange squares and circles indicate the spatial position after the first cycle, Figs.~\ref{fig:Bloch}(a) and \ref{fig:x}. The shift of the position and consequently the changed contour in the parametric space affect the spin behaviour, as will be discussed below.

\begin{figure}[h]
	\centerline{\includegraphics[width=8.5 cm]{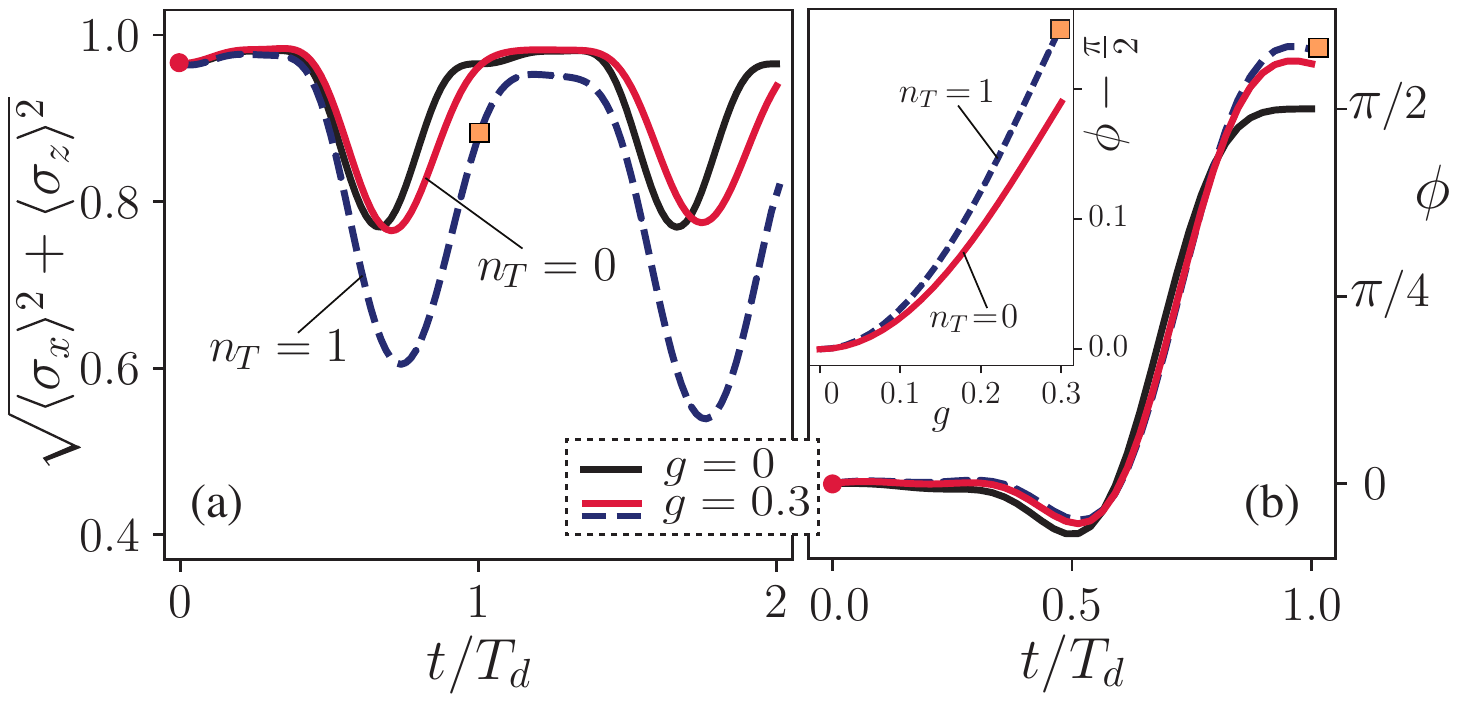}}
	\caption{(a) Time dependence of the spin size $\sqrt{\langle \sigma_z \rangle^2+\langle \sigma_x \rangle^2}$ during two cycles for $g=0$ (black line), $g=0.3$,  $T=0$ (red) and $g=0.3$,  $n_T=1$  (dashed). (b) Angle $\phi$ as a function of time.
Inset: Angle error $\delta \phi=\phi-{\pi \over 2}$ at the final time of one cycle $t=T_d$ as a function of $g$.  In (a) and (b) the red bullets and the orange squares indicate the one cycle starting and final values, correspondingly.}
	\label{fig:spin}
\end{figure}

The spin response of the system is analysed in Fig.~\ref{fig:spin}. First we concentrate on the size of the spin, which for $g=0$ does not depend on the position coordinates regardless of the driving, as can be checked by the application of the unitary transformation Eq.~(\ref{psi}) and a direct evaluation of the spin expectation values. The exact result is 
\begin{equation}\label{eq:sigmana}
\sqrt{\langle \sigma_x \rangle^2+\langle \sigma_z \rangle^2}=e^{- {m^*} \big(a^2_c(t)/\omega_0 +{\dot a}_c^2(t)/\omega_0^3\big)},
\end{equation}
which reduces to $e^{- {m^*} \alpha(t)^2/\omega_0 }$ for the case of slow (adiabatic) driving. In the case of a time independent Rashba coupling this is a known result for Kramers doublets \cite{cadez13}.
Fig.~\ref{fig:spin}(a) shows the size of the spin during the spin-flip protocol in absence of the environment (full black line), the zero temperature result at $g=0.3$ (red line) and the result at a finite temperature $n_T=1$ (dashed line). The orange square indicates the result after the first cycle, as in previous figures.
Effects of the coupling to the bath are more pronounced at an elevated temperature.

The spin rotation, the most relevant property for the qubit manipulation, is shown in Fig.~\ref{fig:spin}(b) for $g=0$ (black) and $g=0.3$, $n_T=0$ (red). Due to the interaction with the bath  the rotation angle is slightly increased with respect to the non-interacting value $\phi={\pi \over 2}$ [the orange square marks the value after one completed cycle]. The deviation $\delta\phi=\phi-{\pi \over 2}$ is additionally presented as a function of $g$ in the inset. Comparing the finite temperature result $n_T=1$ (blue dashed line) with $n_T=0$ (full red) shows that the effect is further increased at finite temperatures. 

\begin{figure}[h]
	\centerline{\includegraphics[width=8 cm]{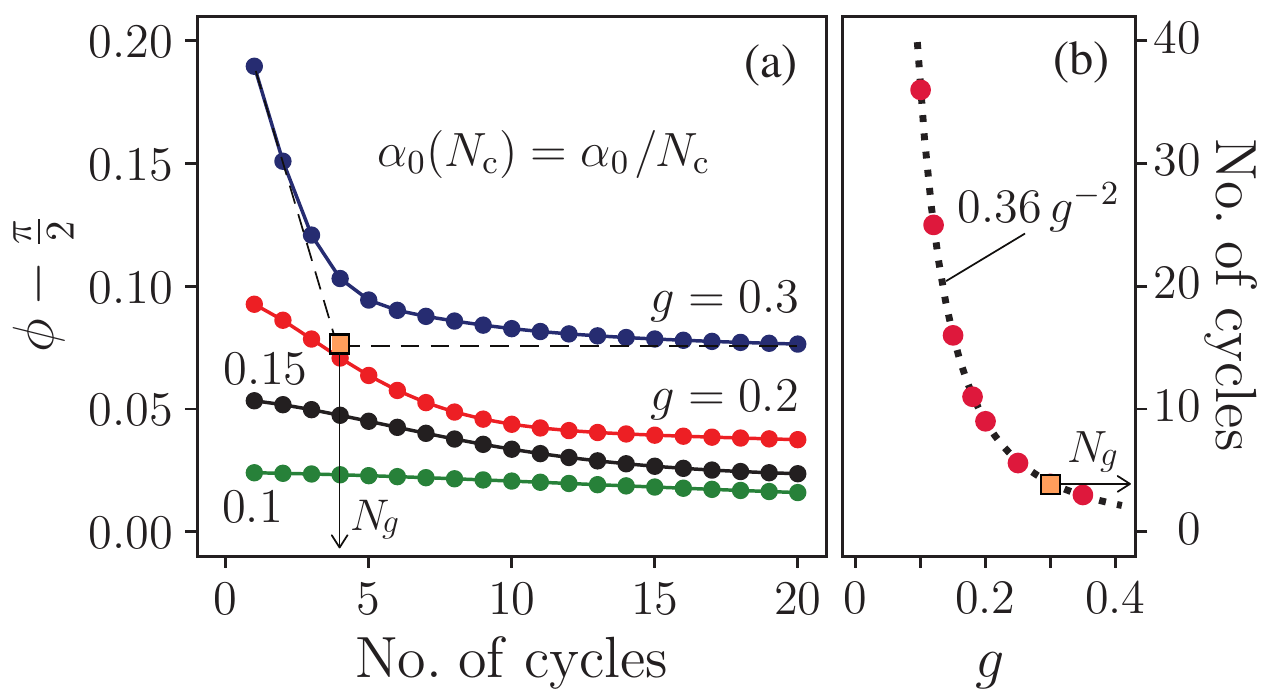}}
	\caption{(a) Deviation of the angle $\phi$ from $\frac{\pi}{2}$ after $N_{\rm c}$ cycles using renormalised Rashba coupling $\frac{\alpha_0}{N_c}$ for various $g$ at $T=0$.  (b) Number of cycles $N_g$  from (a) as a function of $g$. Note the orange symbol relating (a) and (b). The dotted line represents the approximation $N_g\approx 0.36 \,g^{-2}$.}
	\label{fig:angle}
\end{figure}

\subsection{Error analysis and fidelity}

Deviations of $\phi$ from the target value ${\pi \over 2}$ are additionally explored and presented in Fig.~\ref{fig:angle}. Here we performed $N_{\rm c}$ consequent cycles, with renormalised value of the Rashba coupling $\alpha_0(N_{\rm c}) = \alpha_0 / N_{\rm c}$ such that after $N_{\rm c}$ cycles at $g=0$ exactly one half spin-flip is performed. Fig.~\ref{fig:angle}(a) shows that initially the error decreases with an increasing number of cycles. For a large number of cycles, however, the error approaches a constant value. The typical transition number of cycles $N_g$ is for $g=0.3$ indicated by an orange square and an arrow. In Fig.~\ref{fig:angle}(b), $N_g$ is shown as a function of $g$. It clearly exhibits a $g^{-2}$ scaling as expected for the typical relaxation time scale (measured by the number of cycles) for error generating Lamb shift Hamiltonian and Lindblad terms, Eq.~(\ref{eq:lind}).

\begin{figure}[h]
	\centerline{\includegraphics[width=6 cm]{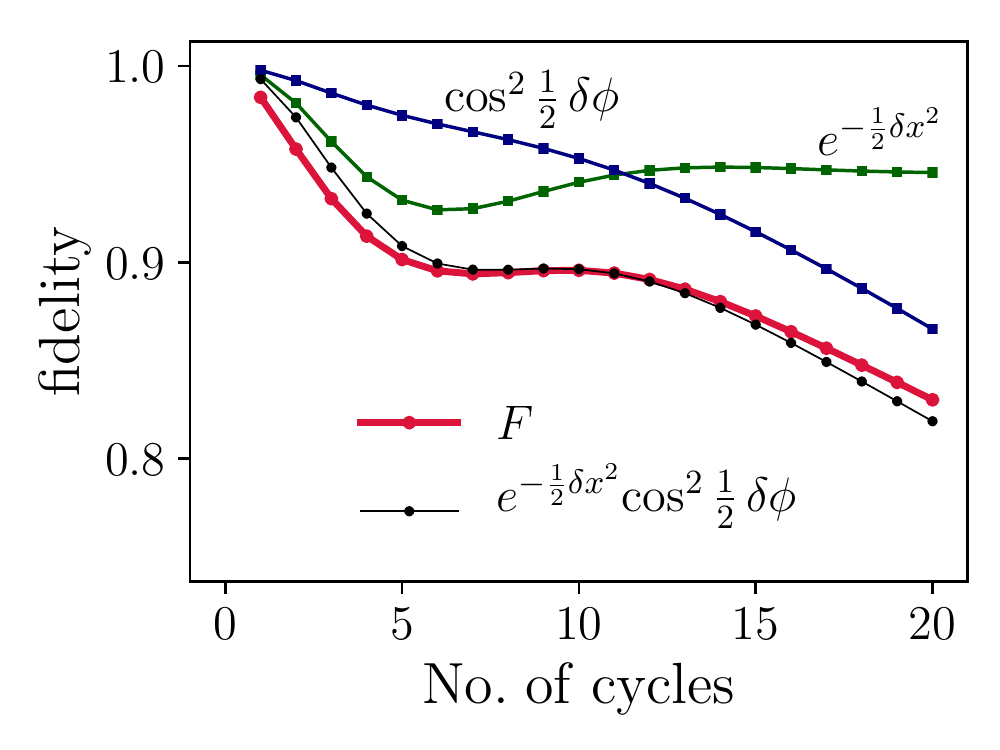}}
	\caption{Uhlmann-Josza fidelity $F$ as a function of the number of cycles for $g=0.2$ and $T=0$ (red). The green squares represent $|\langle\psi_0|\psi_{\delta x}\rangle|^2$, the overlap of two ground states of a harmonic oscillator, one at the origin and the other displaced by $\delta x$. The blue squares represent the overlap $|\langle \chi_0|\chi_{\delta\phi} \rangle |^2$ of two spin states with relative spin angle $\delta \phi$. The black bullets represent the combined overlaps as an estimate of the fidelity.}
	\label{fig:fidelity}
\end{figure}

Let us discuss the effects of the environment also in the framework of fidelity of the spin-qubit transformation. In particular, we consider the Uhlman-Josza fidelity 
\begin{equation}
F=\tr\sqrt{\sqrt{\rho_0}\rho_g\sqrt{\rho_0}},
\end{equation}
where $\rho_0$ and $\rho_g$ represent the density matrix for the non-interacting and the interacting regimes of the model, respectively. At the initial time $t=0$ (or in the absence of interaction) $F=1$, but with increasing time $F$ progressively diminishes due to error generating processes in the Lindblad equation.  Fig.~\ref{fig:fidelity} shows the fidelity calculated at the end of each cycle (red dots) as a function of the number of cycles $N_{\rm c}$ for $g=0.2$ and $T=0$. Here, the Rashba coupling $\alpha_0$ is independent of the number of cycles $N_{\rm c}$. 
 
The structure of $F$ exhibits different short time and long time behaviours. To analyse these behaviours, we make a simple estimate of  fidelity at the end of the driving cycle,
\begin{equation}
F\approx |\langle\psi_0|\psi_{\delta x}\rangle|^2  |\langle\chi_0|\chi_{\delta \phi}\rangle|^2,
\label{F}
\end{equation}
where $|\psi_0\rangle$ is the target final harmonic oscillator ground state and $|\psi_{\delta x}\rangle$ is the ground state of the harmonic oscillator with the potential displaced by $\langle x\rangle-\xi_0=\delta x\xi_0$, giving
\begin{equation}\label{eq:ov1}
 |\langle\psi_0|\psi_{\delta x}\rangle|^2=e^{-{1 \over 2}\delta x^2}.
\end{equation} 
Similarly, $|\chi_{0}\rangle$ is the target final spin state and $|\chi_{\delta \phi}\rangle$ is the spin state with the angle $\delta \phi=\phi-N_{\rm c} {\pi \over 2}$ off from the target state, so that
\begin{equation}\label{eq:ov2}
|\langle\chi_0|\chi_{\delta \phi}\rangle|^2=\cos^2\frac{\delta \phi}{2}.
\end{equation} 
The fidelity (red line and bullets) and its estimate Eq.~\eqref{F} (black), shown in Fig.~\ref{fig:fidelity} for $g=0.2$, behave qualitatively similar. At zero temperature considered here the two overlaps Eqs.~\eqref{eq:ov1} and \eqref{eq:ov2}  represent major sources of the fidelity reduction. The remaining contributions, much more pronounced at finite temperatures (not shown), are mainly due to the fact that the system is not in a pure state and $F$ simply cannot be expressed solely in terms of wave function overlaps. 

The separate curves  $|\langle\psi_0|\psi_{\delta x}\rangle|^2$ (green dots) and $\langle\chi_0|\chi_{\delta \phi}\rangle|^2$ (blue dots) allow us to analyse the short and long-time behaviour of the fidelity. For small number of cycles, $N_{\rm c} \lesssim 10 $, the fidelity is mainly reduced due to the shift of the electron position $\delta x>0$ after a completed cycle. This affects the fidelity due to the reduced overlap of the target spatial wave function and the actual result in the presence of interaction. As discussed before, $\delta x$ exhibits oscillatory behaviour which is damped out (see Fig.~\ref{fig:x}) and $\delta x \to {1 \over 3}$ at larger times $t \gtrsim 10 T_d$, hence the overlap approaches $|\langle\psi_0|\psi_{\delta x}\rangle|^2=e^{-1/18} = 0.95$.

The spin contribution to the fidelity reduction due to the error in the angle of rotation, $\cos^2\frac{\delta \phi}{2}$, is at short times also generated due to oscillations of the orbit $[\langle x \rangle,\alpha]$ and the corresponding deviations from the noninteracting contour ${\cal C}_\alpha$, see Fig.~\ref{fig:Bloch}(a). For $N_{\rm c} \gtrsim 10 $ the orbit in the parametric space progressively relaxes to the steady state contour and then the error $\delta\phi$ increases monotonously, similar to the recent study of adiabatic non-Abelian dephasing \cite{gefen19a,gefen19b}. There are several competing error-generating sources also in the Lamb shift Hamiltonian $H_{LS}$, Eq.~(\ref{HLS}): the spin rotation terms are of the Rashba coupling form $\propto p\, \sigma_y$, a space dependent magnetic field $\propto x\,\sigma_y$ and a constant magnetic field term $\propto \sigma_y$.  Dissipative terms in the Lindblad equation are another important source of the fidelity reduction at larger times. They additionally contribute to spin errors and most importantly, to the size of the spin, shown in Fig.~\ref{fig:spin}(a). At elevated temperatures dephasing effects discussed above amplify due to the increase of coupling factors $S(\omega)$ and $\gamma(\omega)$.

\section{Conclusions}\label{sec:conclusion}

In this paper we have studied the effects of a thermal environment on a non-adiabatic spin-flip protocol. The protocol is based on confining an electron in a harmonic trap and simultaneously manipulating the position of the centre of the trap and the Rashba interaction. 
For arbitrary driving protocols, assuming weak coupling between the system and its environment, the effective dynamics of the system can be obtained in terms of the Lindblad equation \cite{Dann2018}. In the case of periodic driving it reduces to the Floquet-Lindblad equation\cite{BlBu1991,BPfloquet}. However, to obtain the explicit form of the Floquet-Lindblad equation one needs to solve the closed system dynamics, which in the case of periodic driving translates to solving the eigenvalue problem Eqs.~\eqref{eq:Flset}.  
The protocol we are considering is exactly solvable\cite{cadez13}, and the resulting Floquet-Lindblad equation is fully determined by two classical responses Eqs.~\eqref{xcac} to the driving.

Access to the Floquet-Lindblad equation allows us to study the effects of the thermal environment on the driving protocol. The Lindblad equation modifies the free spin-qubit Hamiltonian by introducing low frequencies, an effect called the Lamb shift. Additionally, there are dissipative terms giving rise to spin-dephasing effects as well as thermal activation of the oscillator. 

As an example we consider a specific driving protocol with classical response functions Eq.~\eqref{eq:exresp}. Interestingly, we find that the low frequencies and other terms introduced by the Lamb shift Hamiltonian result in an optimal number of driving cycles to complete the protocol. Figure \ref{fig:angle} displays this result, where depending on the coupling strength to the bath, the optimal number of driving cycles allows us to minimise the error in the final angle of the spin. The Lamb shift Hamiltonian is a time independent shifted harmonic oscillator with spin dependent terms, and the exact solution is a coherent state analogous to Eq.~(\ref{eq:psi}). This allows for an analytical analysis and a deeper understanding of the decoherence dynamics, enabling further possibilities of protocol optimisation at zero and finite temperatures.

The exact study of the considered spin-qubit interacting with the environment can be extended to any driving protocol and any kind of a bath as long as the interaction is weak.

\appendix
\onecolumngrid

\section{Calculation of the Lindblad operators}\label{secapp:ladder}

The time evolution operator Eq.~\eqref{eq:ev} of the free spin-qubit consists of three terms. We calculate  $U^\dagger (t,0) (a+a^\dagger-\sqrt{{2m^*\omega_0}}\xi(t) )U(t,0)$ by first applying $\mathcal{U}^\dagger(t)$, resulting in the expression
\begin{equation}
a+a^\dagger+\sqrt{{2m^* \omega_0 }}\left(x_c(t)-\xi(t)+\frac{\dot{a}_c(t)}{\omega^2_0}\sigma_y\right).
\end{equation}
Applying the time independent harmonic oscillator term $e^{-iH_0 t}$ transforms this into 
\begin{equation}
a e^{-i\omega_0 t}+a^\dagger e^{i\omega_0 t}+\sqrt{{2m^* \omega_0 }}\left(x_c(t)-\xi(t)+\frac{\dot{a}_c(t)}{\omega^2_0}\sigma_y\right).
\end{equation}
Finally, we apply $\mathcal{U}(0)$ and thus obtain
\begin{equation}
\begin{split}
&a e^{-i\omega_0 t}+a^\dagger e^{i\omega_0 t}+\sqrt{{2m^* \omega_0 }}\bigg(x_c(t)-\xi(t)+\frac{\dot{a}_c(t)}{\omega^2_0}\sigma_y-\\
&-x_c(0)\cos(\omega_0t)-\frac{\dot{a}_c(0)}{\omega_0^2}\cos(\omega_0t)\sigma_y+\frac{a_c(0)}{\omega_0}\sin(\omega_0t)\sigma_y-\frac{\dot{x}_c(0)}{\omega_0}\sin(\omega_0t)\bigg).
\end{split}
\end{equation}
Expressing $x_c(t)$ and $a_c(t)$ in terms of their Fourier components $\hat{x}_{c,n}$ and $\hat{a}_{c,n}$ results in the form Eq.~\eqref{eq:ladders}, 
\begin{equation}
\begin{split}
&a e^{-i\omega_0 t}+a^\dagger e^{i\omega_0 t}+\sqrt{{2m^* \omega_0 }}\sum_{\substack{n\in \mathbb{Z}\\n\ne\pm n_d}}
\left(\hat{x}_{c,n}-\hat{\xi}_n+\frac{\hat{\dot{a}}_{c,n}}{\omega^2_0}\sigma_y\right)e^{-in \omega_d t}+\\
&+\sqrt{{\frac{m^* \omega_0}{2} }}\left(-x_c(0)-i\frac{\dot{x}_c(0)}{\omega_0}-\frac{\dot{a}_c(0)}{\omega_0^2}\sigma_y+i\frac{a_c(0)}{\omega_0}\sigma_y\right){e^{-i\omega_0 t}}+\\
&+\sqrt{{\frac{m^* \omega_0}{2} }}\left(-x_c(0)+i\frac{\dot{x}_c(0)}{\omega_0}-\frac{\dot{a}_c(0)}{\omega_0^2}\sigma_y-i\frac{a_c(0)}{\omega_0}\sigma_y\right){e^{i\omega_0 t}},
\end{split}
\end{equation}
from which we read the jump operators of Eq.~\eqref{eq:ladder1}. The Fourier components $\hat {f}_n$ are defined by $f(t)=\sum_{n\in\mathbb{Z}} {\hat f}_n e^{-i n \omega_d t}$  and are interconnected by useful relations
$\hat{\dot{a}}_{c,n}=-i {n \over n_d} \omega_0 \hat{a}_{c,n}$ and, for $n\ne\pm n_d$,    $\hat{a}_{c,n}={n_d^2 /( {n_d^2-n^2})}\hat{\alpha}_n$ and
$\hat{x}_{c,n}-\hat{\xi}_n={n^2 /( {n_d^2-n^2})}\hat{\xi}_n$.

\section{Lamb shift Hamiltonian}\label{secapp:ladder1}

The Lindblad equation is invariant under inhomogeneous transformations
\begin{subequations}\label{B1}
\begin{align}
	& A_n\rightarrow \bar{A}_n=A_n+z_n,\\
	& H_{LS}\rightarrow \bar{H}_{LS}=H_{LS} +\frac{1}{2i}\sum_{n\in\mathbb{Z}} \gamma(n\omega_d)\left(z_n^\ast A_n-z_nA_n^\dagger\right)+c
\end{align}
\end{subequations}
where $z_n\in \mathbb{C}$ and $c\in \mathbb{R}$. The  Lindblad operators Eq.~\eqref{eq:ladder1} consist of two parts, one proportional to $\sigma_y$, $a$ and $a^\dagger$ and the other proportional to the identity. The latter can be eliminated using the above transformation, leading to Eq.~\eqref{eq:Lindf} with the transformed Lamb shift Hamiltonian 
\begin{equation}\label{hlsbar}
\bar{H}_{LS}=H_{LS}
-\frac{1}{2i}g^2\omega_0\sqrt{2 m^* \omega_0}\left(\big(\hat{x}_{c,-n_d}-\hat{\xi}_{-n_d}-\frac{1}{2}x_c(0)+i{\dot{x}_c(0) \over 2\omega_0}\big)A_{n_d} - \mathrm{h.c.}\right)+\bar{B}\sigma_y
\end{equation}
where 
\begin{equation}
\bar{B}=2g^2 m^*\sum_{\substack{n\in \mathbb{N}\\ n\ne n_d}}
{n^5 \omega_0\over n_d\left( {n_d^2}-n^2\right)^2} \mathrm{Re}\left\{ \hat{\alpha}_n \hat{\xi}_n^\ast\right\}.
\end{equation}

In the particular case of an even periodic driving function $\xi(t)$ and an odd periodic driving function $\alpha(t)$, i.e., when $\xi(-t)=\xi(t)$ and $\alpha(-t)-\alpha(0)=-\left(\alpha(t)-\alpha(0)\right)$ [as is the case in the example studied in Section~\ref{sec:example}], $x_c(-t)=x_c(t)$ and $\dot{a}_c(-t)=\dot{a}_c(t)$ and the Lamb shift Hamiltonian Eq.~(\ref{Lambshift}) simplifies. It can be represented as a shifted harmonic oscillator in the presence of the Rashba interaction and an inhomogeneous magnetic field, 
\begin{subequations}\label{HLS}
\begin{eqnarray}
H_{LS}&=&\frac{p^2}{2m_{LS}} + \frac{m_{LS}\omega_{LS}^2}{2}(x-x_{LS})^2 
+\big(\alpha_{LS}p +b_{LS}x+B_{LS}\big)\sigma_y,\\
B_{LS}&=&-b_{LS} x_{LS}-
4\zeta m_{LS}\sum_{\substack{n\in \mathbb{Z}\\n\ne\pm n_d}}
{n_d n^3 \over ({n_d^2-n^2})^2}S(n\omega_d)
\;i\hat{{\alpha}}_n\hat{\xi}_n.
\end{eqnarray}
\end{subequations}
Here $\zeta=\left(S(\omega_0)+S(-\omega_0)\right)/\omega_0$, $m_{LS}=\zeta^{-1}m^*$, $\omega_{LS}=\zeta \omega_0$, $x_{LS}=x_c(0)-2\big(\hat{x}_{c,n_d}-\hat{\xi}_{n_d}\big)$, $\alpha_{LS}=\zeta a_c(0)$, $b_{LS}=\zeta^2m_{LS}\big(2\hat{\dot{a}}_{c,n_d}-\dot{a}_c(0)\big)$
and, at $T=0$,
\begin{equation}\label{Spm}
 \zeta=-\frac{g^2}{2\pi}\left(\frac{\omega_c (\omega_c+2\omega_0)}{2\omega_0^2}+\log \frac{\left|\omega_c-\omega_0\right|}{\omega_0}\right),
 \end{equation}
which is zero at $\omega_c=1.16\,\omega_0$ and at $\omega_c=2\omega_0$, as used throughout the paper, $\zeta=-{1 \over \pi}g^2$. Note that the resonant frequency components of driving, $\hat{\xi}_{\pm n_d}$, should vanish if the steady state regime of driving and response is to be studied. Note also that $\hat{\xi}_{\pm n_d}=0$ does not imply $\hat{x}_{c,\pm n_d}=0$, respectively. Applying the transformation Eqs.~(\ref{B1}) results in $\bar{B}=0$ and 
\begin{equation}
\bar{H}_{LS}=H_{LS}+\frac{1}{2}g^2\omega_0x_{LS}\left(p+m_{LS}\alpha_{LS}\sigma_y\right).
\end{equation}

\twocolumngrid
\bibliography{bibliography_qubit_list}

\end{document}